\documentclass[epjCONF]{svjour}
\usepackage{graphics}
\usepackage[varg]{txfonts} 
\usepackage[latin1]{inputenc}
\usepackage{units}
\usepackage{url}

\usepackage{color}
\usepackage{epic,eepic,pstricks}

\newcommand*\patchAmsMathEnvironmentForLineno[1]{%
  \expandafter\let\csname old#1\expandafter\endcsname\csname #1\endcsname
  \expandafter\let\csname oldend#1\expandafter\endcsname\csname end#1\endcsname
  \renewenvironment{#1}%
     {\linenomath\csname old#1\endcsname}%
     {\csname oldend#1\endcsname\endlinenomath}}%

\session-title{UHECR 2012}
\begin{document}
\title{ Measurement of the energy spectrum of cosmic rays at the
  highest energies using data from Pierre Auger Observatory}
\author{Ioana~C.~Mari\c{s}
  \inst{1}\fnmsep\thanks{\email{Ioana.Maris@lpnhe.in2p3.fr}} for the
  Pierre Auger Collaboration \inst{2}}
\institute{LPNHE, 4 place Jussieu, 75252 Paris CEDEX 05, France \and
  Observatorio Pierre Auger, Av. San Mart{\'\i}n Norte 304, 5613
  Malarg\"ue, Argentina\\ (Full author list:
  \url{http://www.auger.org/archive/authors_2012_06.html})}
\abstract{ We report a measurement of the flux of cosmic rays with
  unprecedented precision and statistics using data from the Pierre
  Auger Observatory. Based on fluorescence observations in coincidence
  with at least one station of the surface detector we derive a
  spectrum for energies above $10^{18}$ eV. We also report on the
  energy spectra obtained with the surface detector array. The
  spectral features are presented in detail and the impact of
  systematic uncertainties on these features are addressed.  }
\maketitle
\section{Introduction}
\label{intro}

The propagation and  the origin of cosmic rays influence the number
of particles that enter the Earth's atmosphere. The evolution of the
flux with energy exhibits at the highest energies two features: the
ankle, a flattening of the flux at about \unit[$4 \times 10^{18}$]{eV},
and a strong suppression above \unit[$5\times 10^{19}$]{eV}. Currently
it is not clear if these features are caused by the interaction of the
particles with the cosmic background, or if they are hints of changes
in the acceleration origin or
power~\cite{Berezinsky:2005cq,Hillas:2005cs,Wibig:2004ye,Allard:2008gj}. A
precise measurement of the flux of cosmic rays is an important
ingredient in discriminating between theoretical models.

The Pierre Auger Observatory is designed as an hybrid detector: a
surface detector (SD) with more than 1600 water-Cherenkov stations
placed on a triangular grid covering $\unit[3000]{km^2}$ and a
fluorescence detector (FD) made of 27 optical telescopes grouped in
five buildings, observing the atmosphere above the array. The detailed
description of the Observatory can be found in
~\cite{EngArray,SDDet,FDDet}. The hybrid design allows for the energy
calibration of the SD data and for obtaining the energy spectrum over
two decades in energy. We present the measurements of the flux of
cosmic rays above \unit[$10^{18}$]{eV}.

\section{Air-showers reconstruction}
The FD observes the fluorescence and Cherenkov light produced by the
secondary particles from the air-showers in their passage through the
atmosphere. The arrival direction is obtained from the timing
information of the signal in individual triggered pixels and from at
least one station of the SD. This set of events are referred to as
{\em hybrid} events. From the reconstruction of the longitudinal
profiles~\cite{michael} the energy of the primary particle is
inferred in an almost calorimetric way. The invisible energy, the
fraction of energy carried by the muons and neutrinos, is obtained
from parametrization of simulations assuming a mixed composition and
QGSJet01 as hadronic interaction model. It is less than 10\% above
$\unit[10^{18}]{eV}$. The energy resolution is
$7.6\%$~\cite{FrancescoICRC}.

\begin{figure}\sidecaption
  \resizebox{0.41\columnwidth}{!}{
    \includegraphics{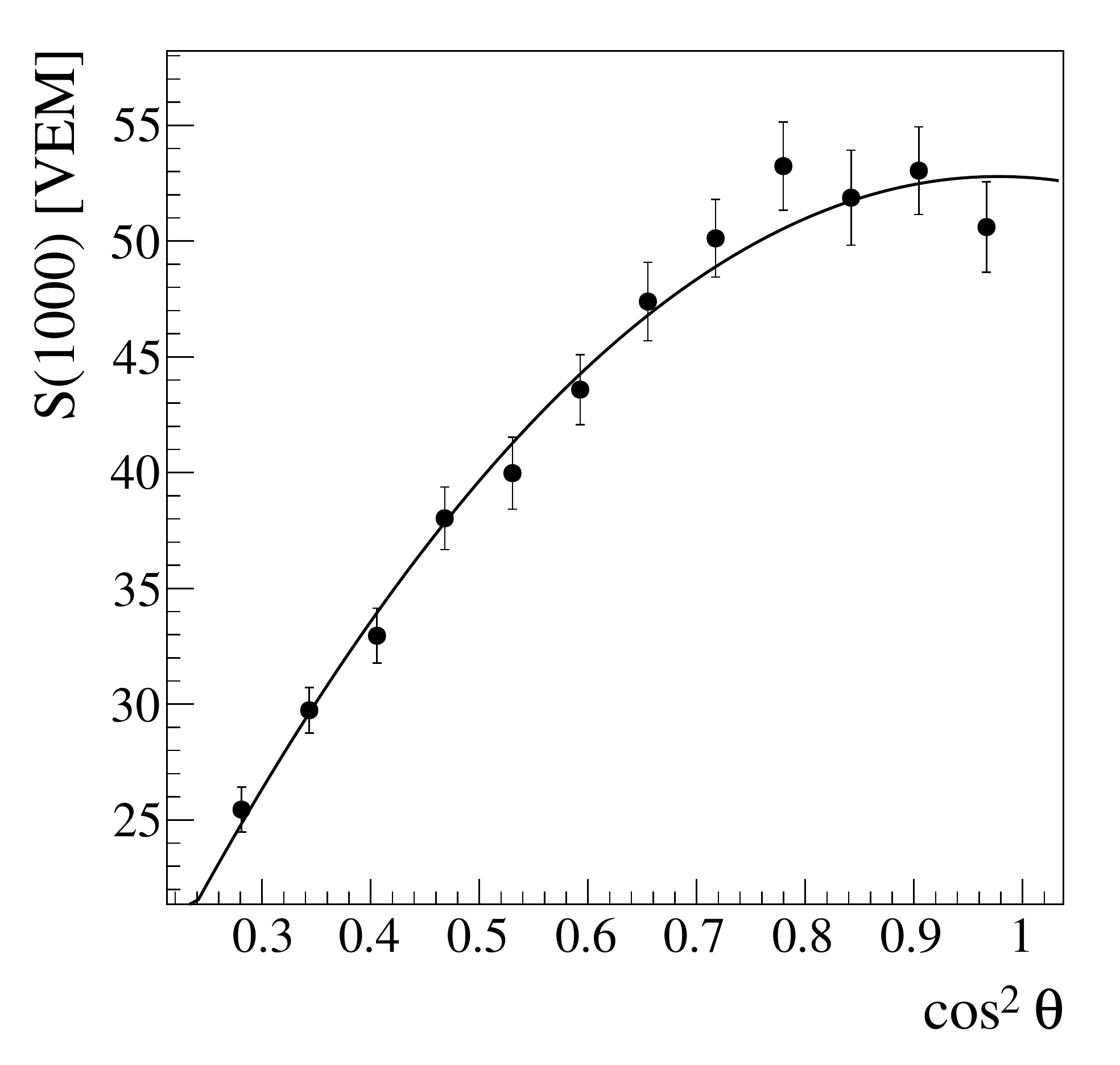}
  }
  \resizebox{0.51\columnwidth}{!}{
    \includegraphics{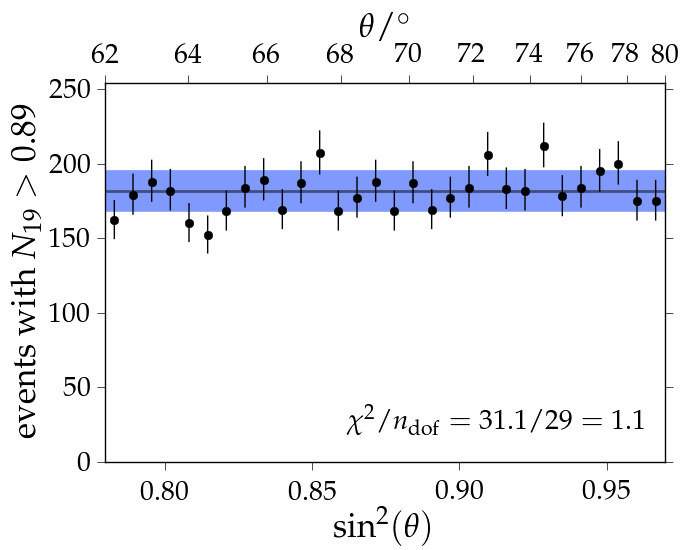}
  }
  \caption{ (left) S(1000) as a function of the
    zenith angle. (right) The number of events above $N_{19} = 0.89$ as a
    function of the zenith angle.}
  \label{fig:att}
\end{figure}

The signal from the particles reaching the ground is sampled with the
water-Cherenkov detectors. The content of the air-showers at this level
depends on the amount of matter traversed by the cascade. For
{\em vertical} events (zenith angle smaller than 60 degrees) the signal
contains contributions from both the electromagnetic and muonic
components of the air-showers. At angles above 62 degrees ({\em inclined}
events), the electromagnetic component is suppressed. This
effect leads to different reconstruction methods for vertical and
inclined events. In the first case the signal at \unit[1000]{m} from
the air-shower axis, S(1000), is interpolated from a lateral distribution
function. S(1000) is corrected for the attenuation in the atmosphere
through a constant intensity method: an air-shower with a zenith angle
of 60 degrees produces a signal smaller by a factor of two then if it
would have had an arrival direction of zero degrees (see
Fig.~\ref{fig:att} (left)). The energy estimator for the vertical
events is $S_{38}$, the equivalent signal for a zenith angle of 38
degrees.~\cite{RobertoICRC,SpectrumSD}.  The reconstruction of the
inclined events is based on a detailed modeling of the two dimensional
muon density on the ground~\cite{InclinedReco}. The energy
estimator, $N_{19}$ is obtained by taking as a reference the
normalization of the footprint of proton initiated air-showers with an
energy of $\unit[10^{19}]{eV}$ and it is independent of the zenith
angle(Fig.~\ref{fig:att}(right)).

\begin{figure}
  \centering
  \resizebox{0.42\columnwidth}{!}{
    \includegraphics{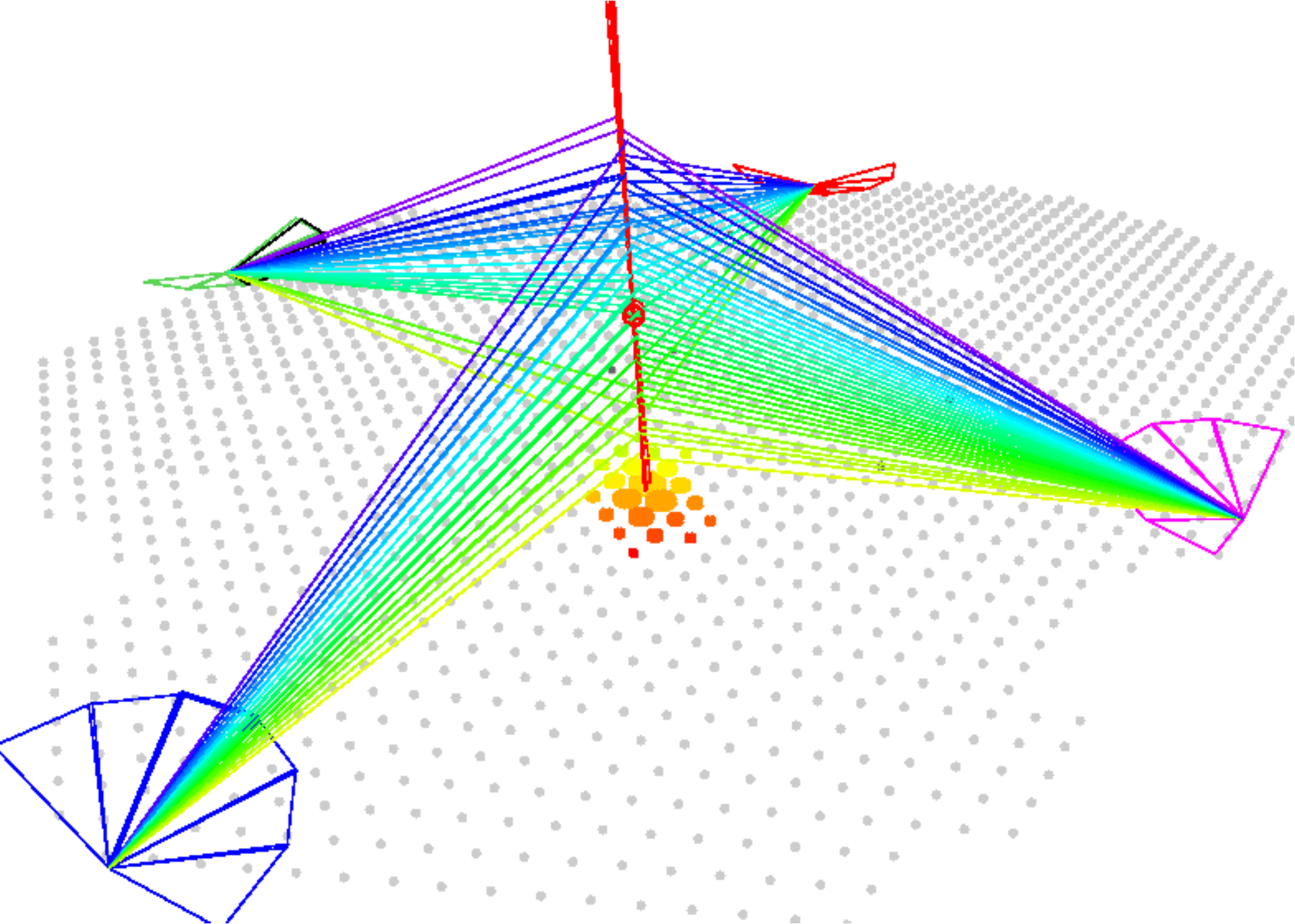}
  }
  \resizebox{0.42\columnwidth}{!}{
    \includegraphics{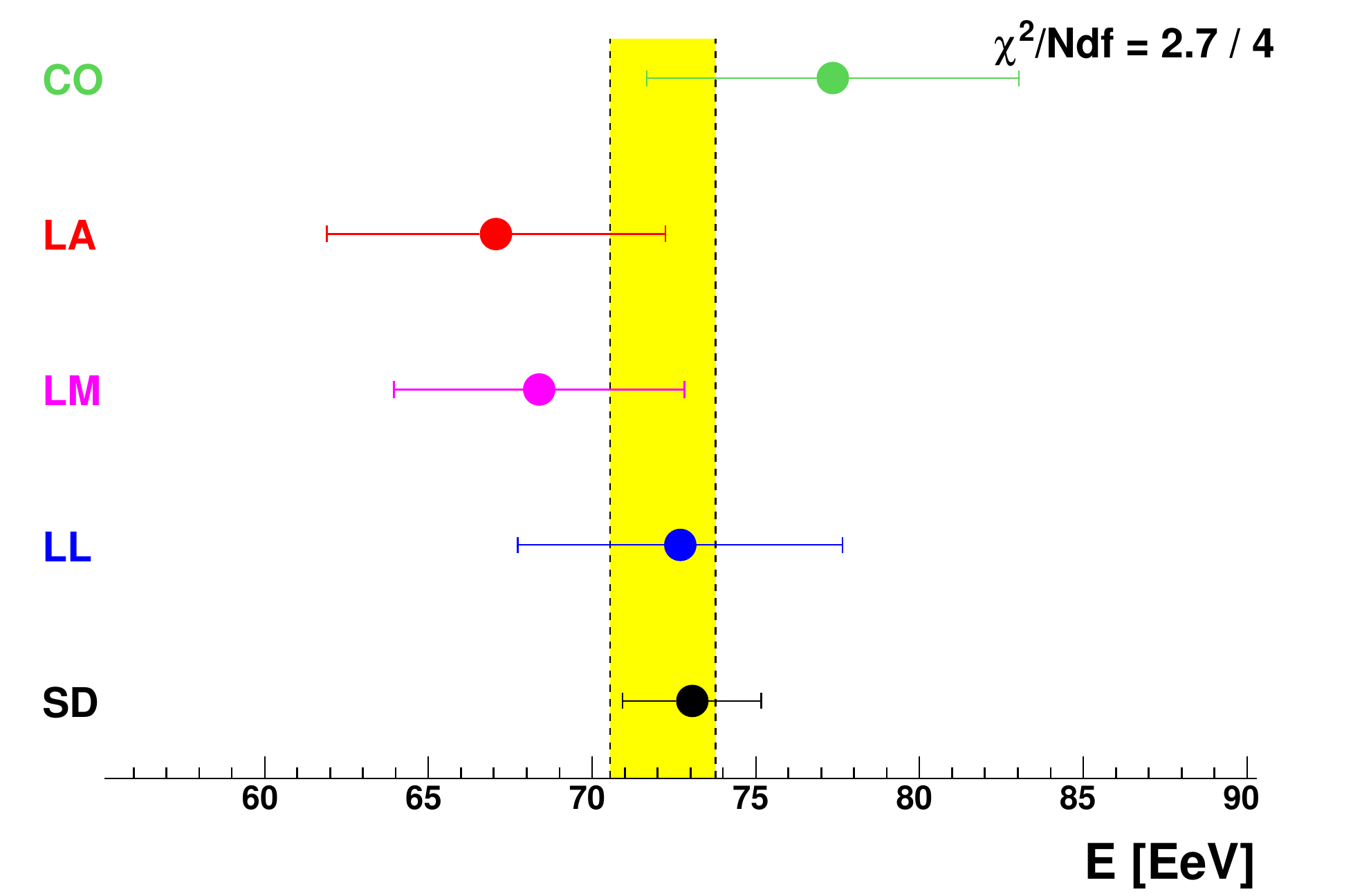}
  }
  \caption{Example of an event with an energy of $\unit[7\times
      10^{19}]{eV}$ falling in the middle of the surface detector and
    observed with at least one telescope from each of the FD buildings
    (left); (right) the independent energy assignments from the four
    eyes (CO, LA, LM, LL) and from the surface detector (SD).}
  \label{fig:evExample}
\end{figure}

An event example with an energy of $\unit[7\times 10^{19}]{eV}$ is
shown in Fig.~\ref{fig:evExample}. The air-shower has triggered at
least one telescope in each FD buildings and the surface detector. The
energies reconstructed independently with the data from the four eyes
are in a very good agreement, even if the air-shower is more than
\unit[20]{km} away. The energy measured from the SD data agrees with
the FD reconstructions. The details of how we measure the energy with
the SD are given next.

\paragraph{Energy calibration and resolution}

\begin{figure}\sidecaption
  \centering
  \resizebox{0.5\columnwidth}{!}{
    \includegraphics{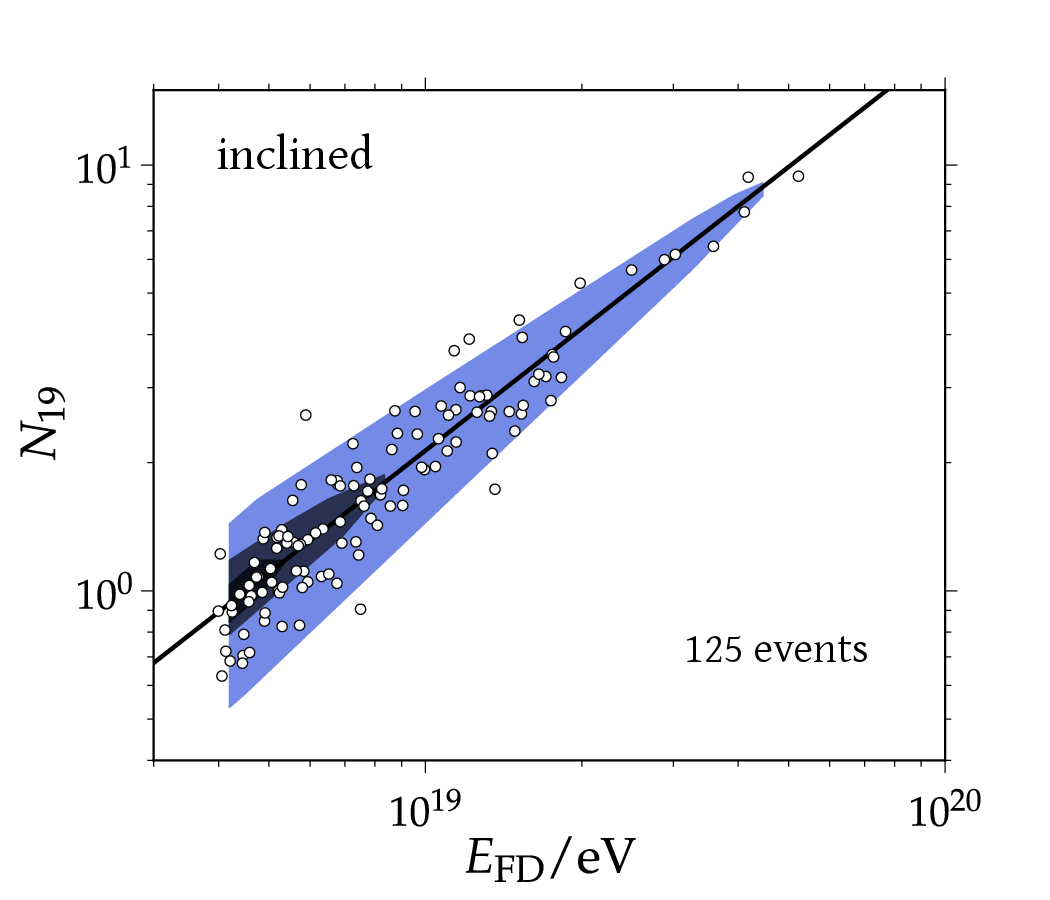}
  }
  \resizebox{0.42\columnwidth}{!}{
    \includegraphics{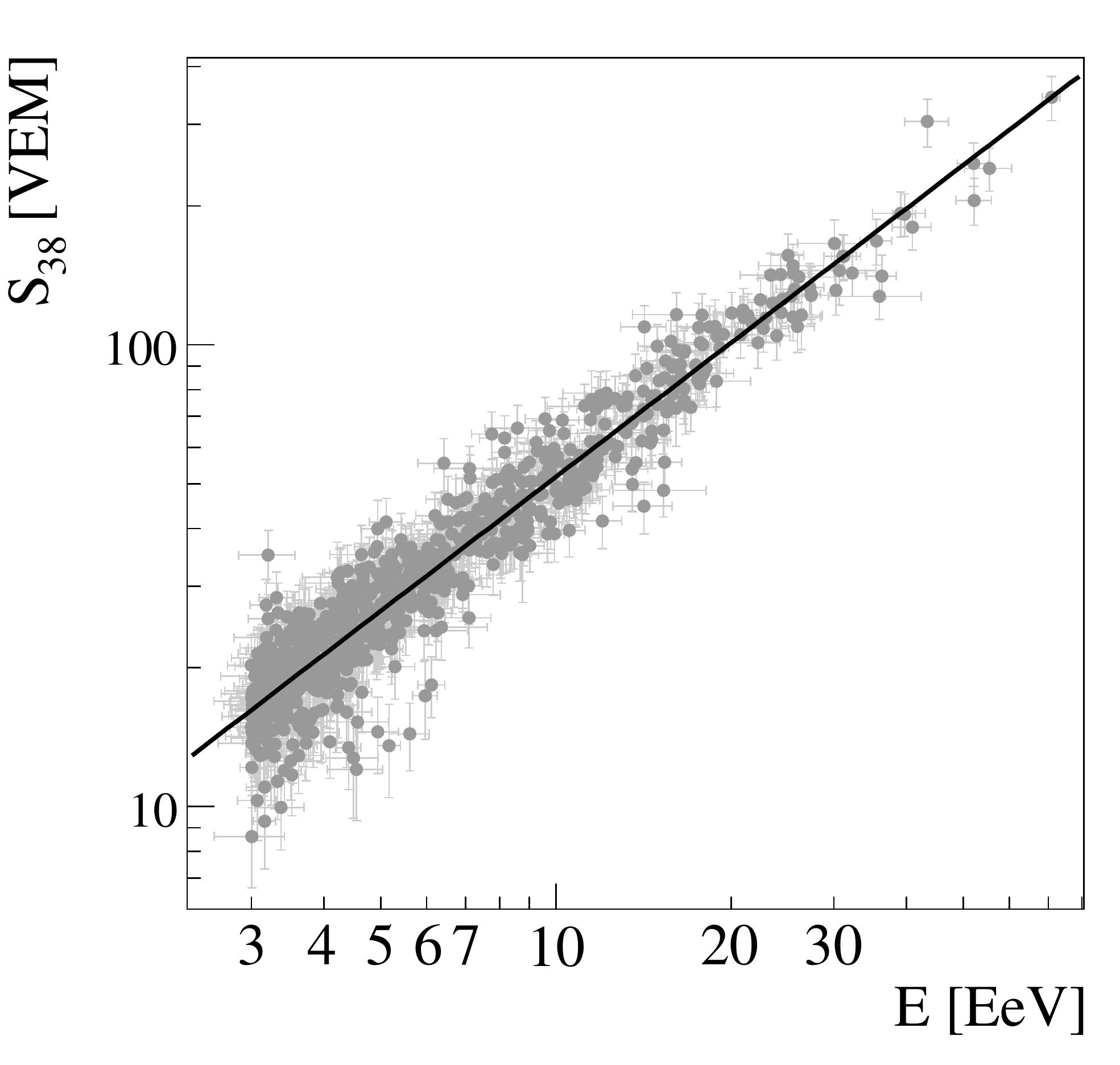}
  }
  \caption{The energy calibrations curves for the inclined events
    (left) and for the vertical events (right). The lines represent
    the best fit to data. }
  \label{fig:enCalib}
\end{figure}

To convert $N_{19}$ and $S_{38}$ to energy we use golden hybrid
events: a sub-sample of the hybrid events which are independently
triggered and reconstructed with the SD. We are selecting high quality
longitudinal profiles, observed during a time period with clear
atmosphere~\cite{Abraham:2010pf}, with a cloud coverage of less than 25\%. Deeply
penetrating air-showers would produce a larger signal on the ground
that shallow ones. Thus, in order to reduce a possible mass
composition bias we apply strict cuts on the fiducial field of view
which assure an equal hybrid trigger probability for different
primaries.

\begin{figure}\sidecaption
  \centering
  \resizebox{\columnwidth}{!}{
    \includegraphics{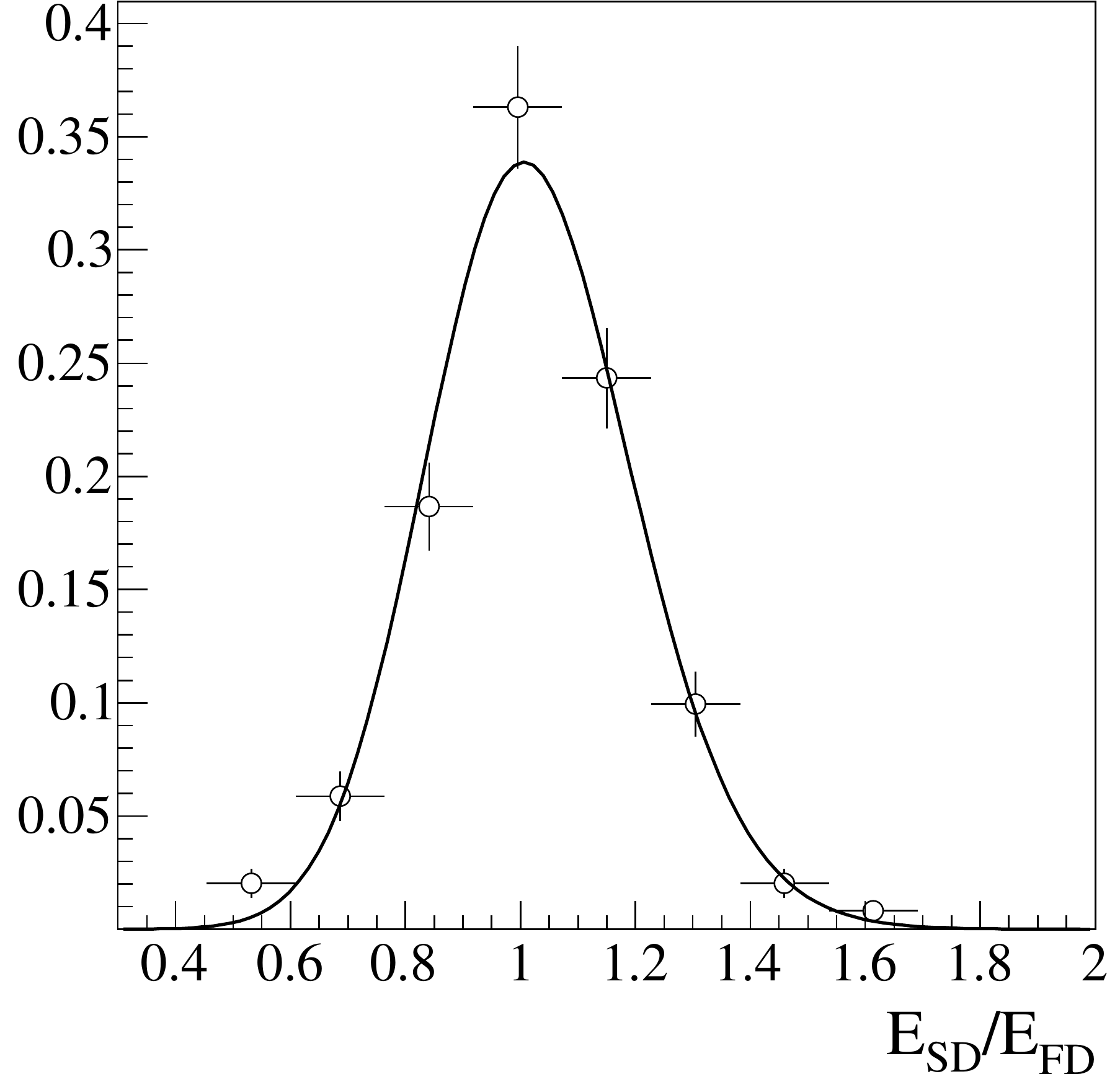}
    \includegraphics{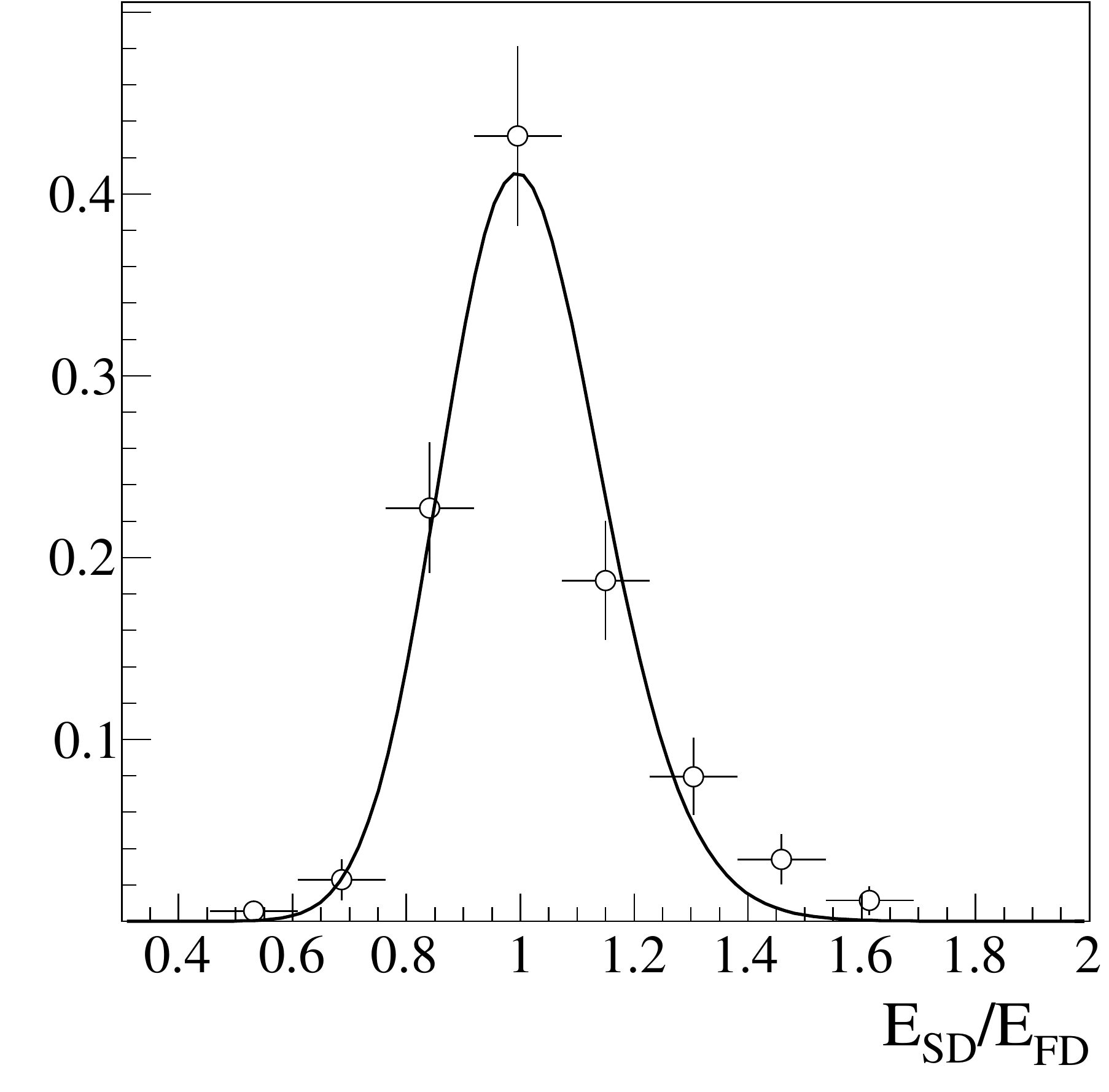}
    \includegraphics{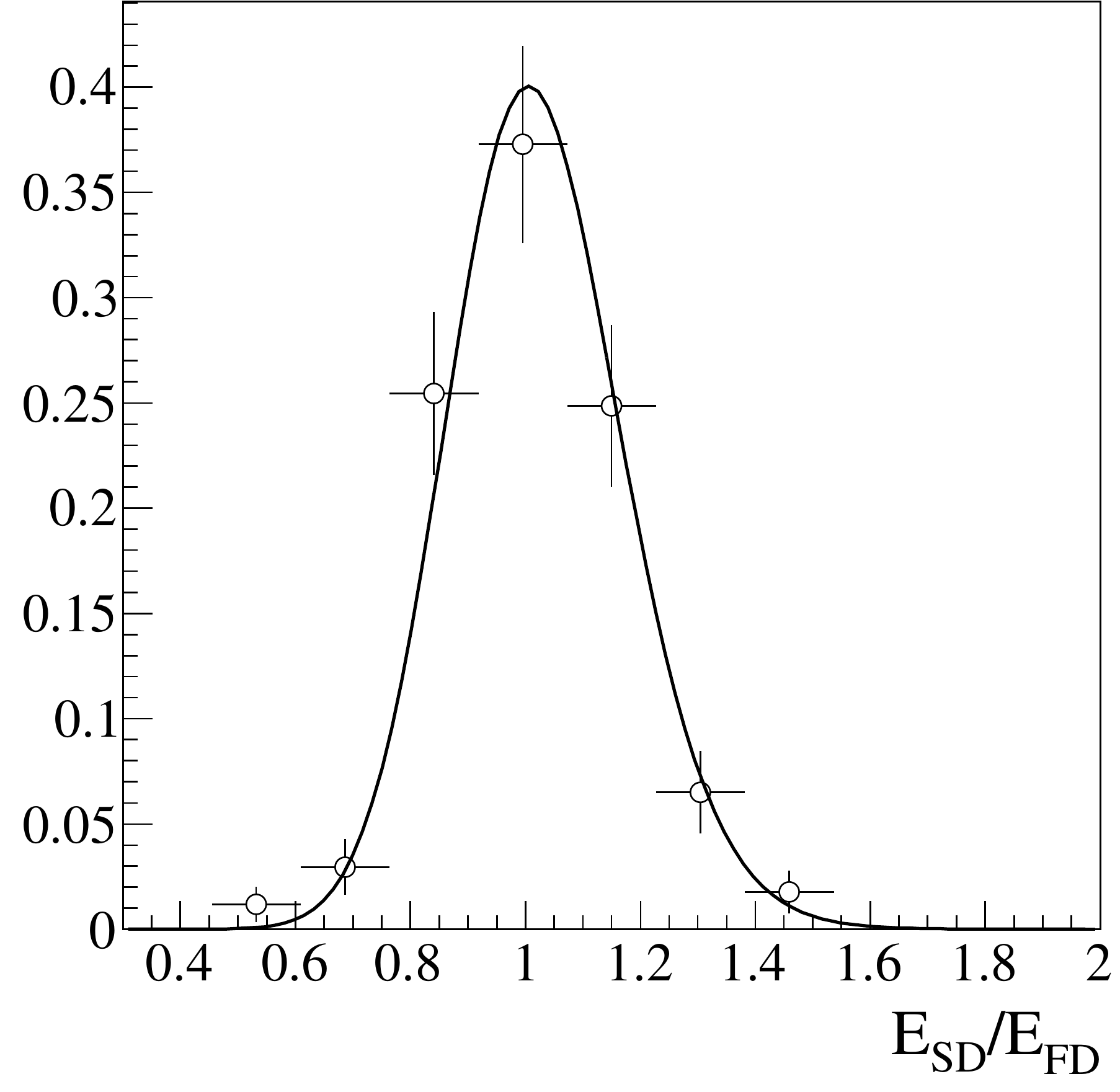}
  }
  \caption{The ratio between the energies reconstructed with the
    hybrid and the surface detector data for different energy
    intervals: between $\unit[3\times 10^{18}]{eV}$ and $\unit[6\times
      10^{18}]{eV}$ (left) between $\unit[6\times 10^{18}]{eV}$ and
    $\unit[10^{19}]{eV}$ (center) and above $\unit[10^{19}]{eV}$ (right).}
  \label{fig:enRes}
\end{figure}

The correlation functions are obtained by maximizing the
likelihoods and are in both cases almost linear~\cite{RobertoICRC,HansICRC}:
\begin{equation}
E_{\rm inclined} = \unit[(4.69\pm0.09)\times 10^{18}]{eV}\cdot N_{19}^{(1.05\pm0.02)},
\hspace*{2ex}
E_{\rm vertical} = \unit[(1.68\pm0.05)\times 10^{17}]{eV}\cdot S_{38}^{(1.035\pm0.009)}
\end{equation}
The energy calibrations, shown in Fig.~\ref{fig:enCalib}, induce a
systematic uncertainty on the energy due to the limited statistics:
125 events for the inclined data and 839 for the vertical data. Namely
7\% at $\unit[10^{19}]{eV}$ and 12\% at $\unit[10^{20}]{eV}$ for the
vertical events and 13\% at $\unit[10^{19}]{eV}$ and more than 50\% at
$\unit[10^{20}]{eV}$ for the inclined events. In the future with more
accumulated data, these uncertainties will be reduced.  Assuming a
constant energy resolution for the hybrid energy of $\approx 8\%$, the
energy resolution for the vertical data set can be obtained from the
golden hybrid events (Fig.~\ref{fig:enRes}). The resulting SD energy
resolution decreases from $\sigma_{E}/E_{\rm vertical} = (15.8\pm 0.9)\%$
below $\unit[6\times 10^{18}]{eV}$ to $(12.0 \pm 1.0)\%$ for $E_{\rm
  SD} >\unit[10^{19}]{eV}$, the reconstruction uncertainties being
important at $\unit[3\times 10^{18}]{eV}$, while at the highest
energies the shower-to-shower fluctuations of S(1000) are dominant.

\begin{figure}\sidecaption
  \centering
  \resizebox{\columnwidth}{!}{
      \includegraphics{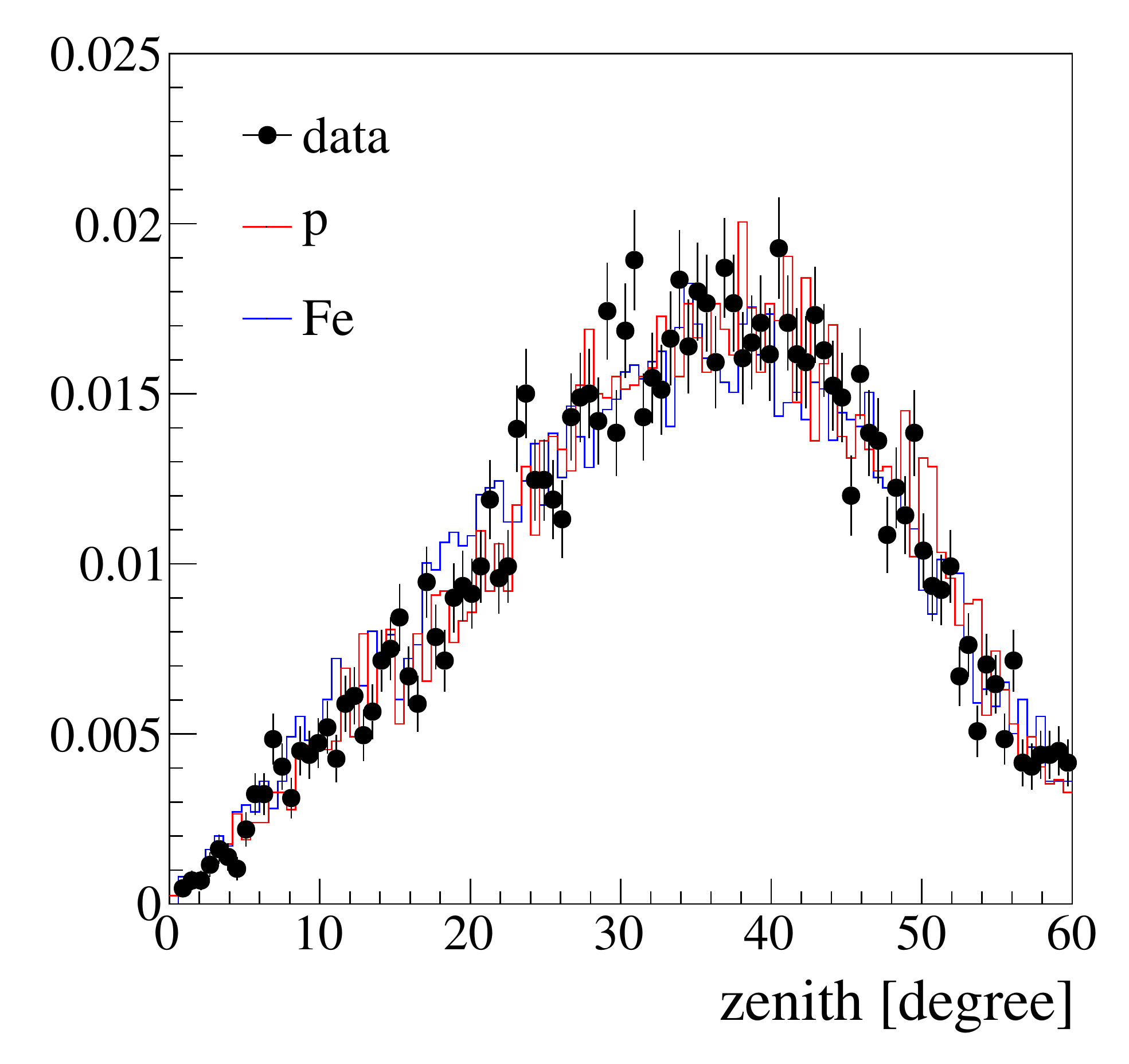}
      \includegraphics{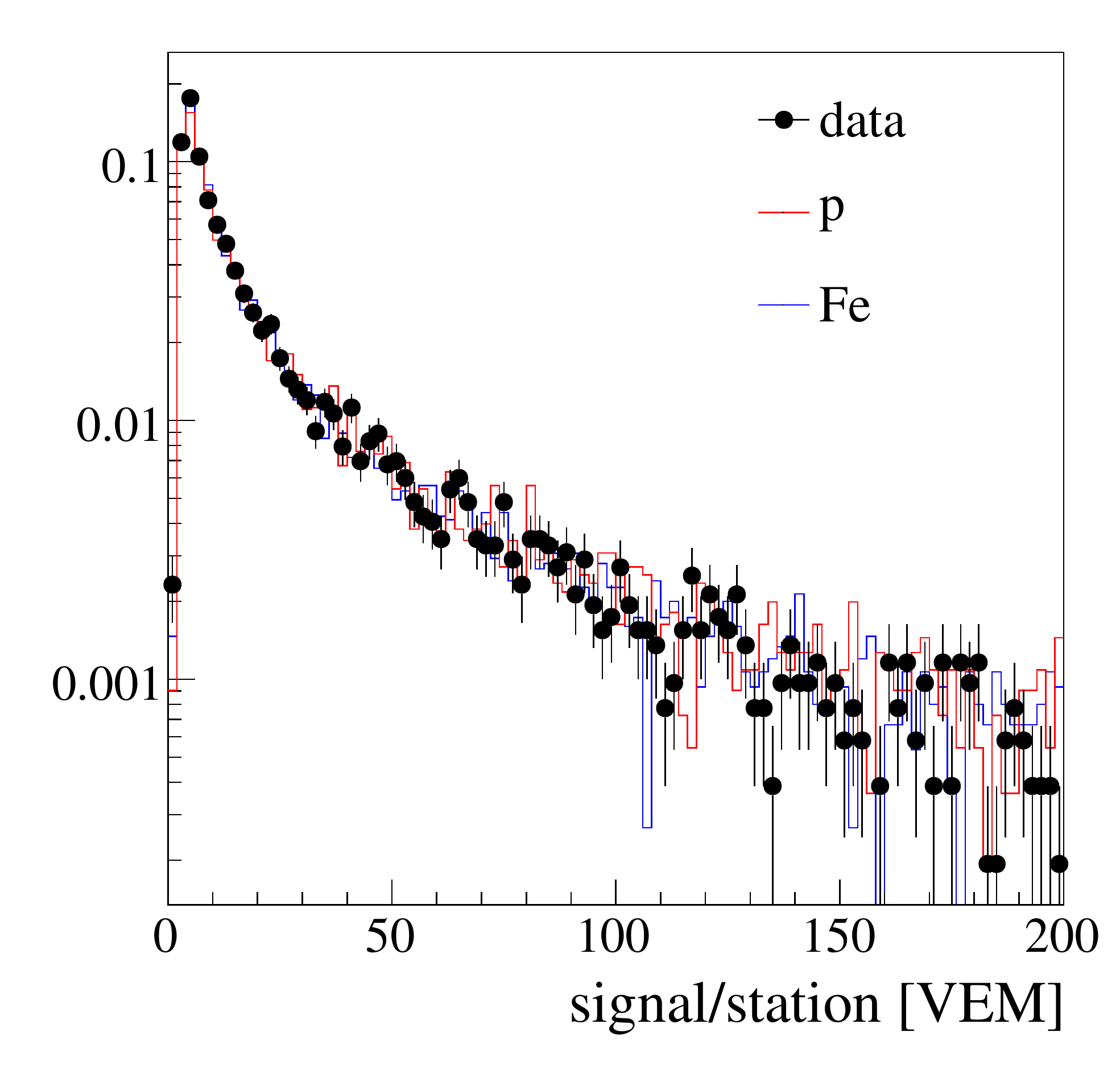}
      \includegraphics{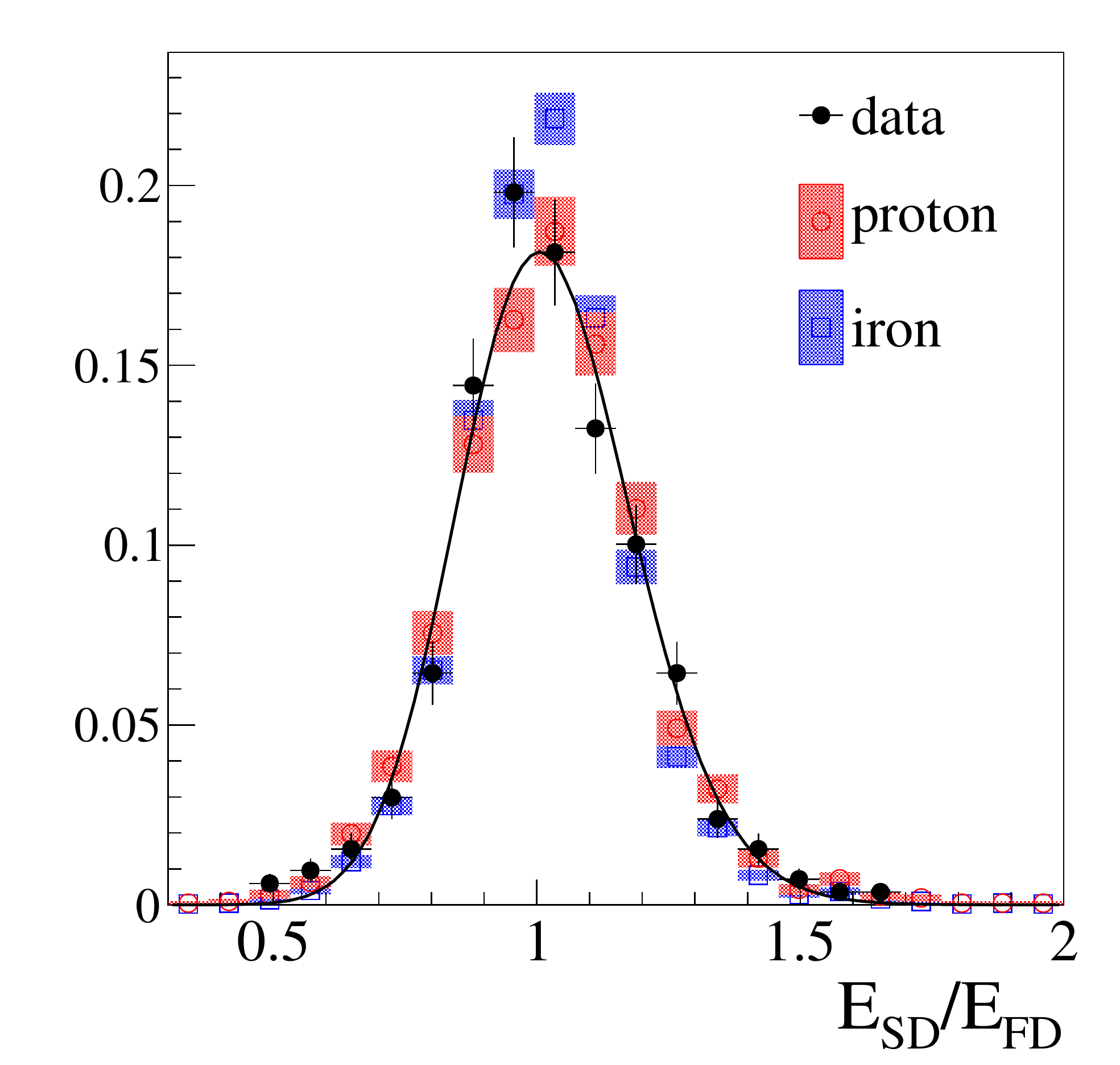}
    }
  \caption{Comparison between data and MC. The distributions are shown
    for QGSJet II.03 for proton and iron. Simulations have been
    rescaled to take into account the missing number of muons.  (left)
    Normalized distribution of the number of events as a function of
    the zenith angle; (center) Normalized distribution of the signal
    in individual stations for all the stations in the recorded and
    simulated events; (right) Normalized distribution of the FD and SD
    energy ratio. }
  \label{fig:dataMC}
\end{figure}

The three measurements with inclined, vertical and hybrid events
inherit the same systematic uncertainty on the energy from the
fluorescence detector energy assignment. The determination of the
absolute value of the fluorescence yield contributes with 14\%, the
reconstruction of the energy with 4\% due to the invisible energy
correction, and 10\% from the reconstruction of the longitudinal
profiles, while the atmospheric effects add another 8\%. The total
systematic uncertainty of 22\%, includes as well the calibration of
the detector, and the propagated uncertainties of the aerosol optical
depth measurements~\cite{RobertoICRC}.

\section{Energy spectrum}

The energy resolution of the surface detector is energy dependent and
thus it distorts the distribution of events as a function of energy by
bin-to-bin migrations.  To obtain the true number of events a
forward-folding procedure is used. The migration matrices are built
from air-showers and detector simulations by using
CORSIKA~\cite{corsika} with QGSJet II.03~\cite{qgsjetII} as hadronic
interaction model.  Even if the number of muons is underestimated in
the simulations~\cite{Muons}, after an energy rescaling such that for
the same true $S_{38}$ we obtain the same energy in data and
simulations, the distributions in the main shower observables are
reproduced by the simulations: in Fig.~\ref{fig:dataMC} are depicted the
normalized distribution for events with an energy above $\unit[3\times
  10^{18}]{eV}$ of the zenith angle, of the signal per station
including all stations in the events and of the ratio of the energy
determinations with the FD and SD, for proton and iron primaries.

\begin{figure}\sidecaption
  \resizebox{0.62\columnwidth}{!}{
    \includegraphics{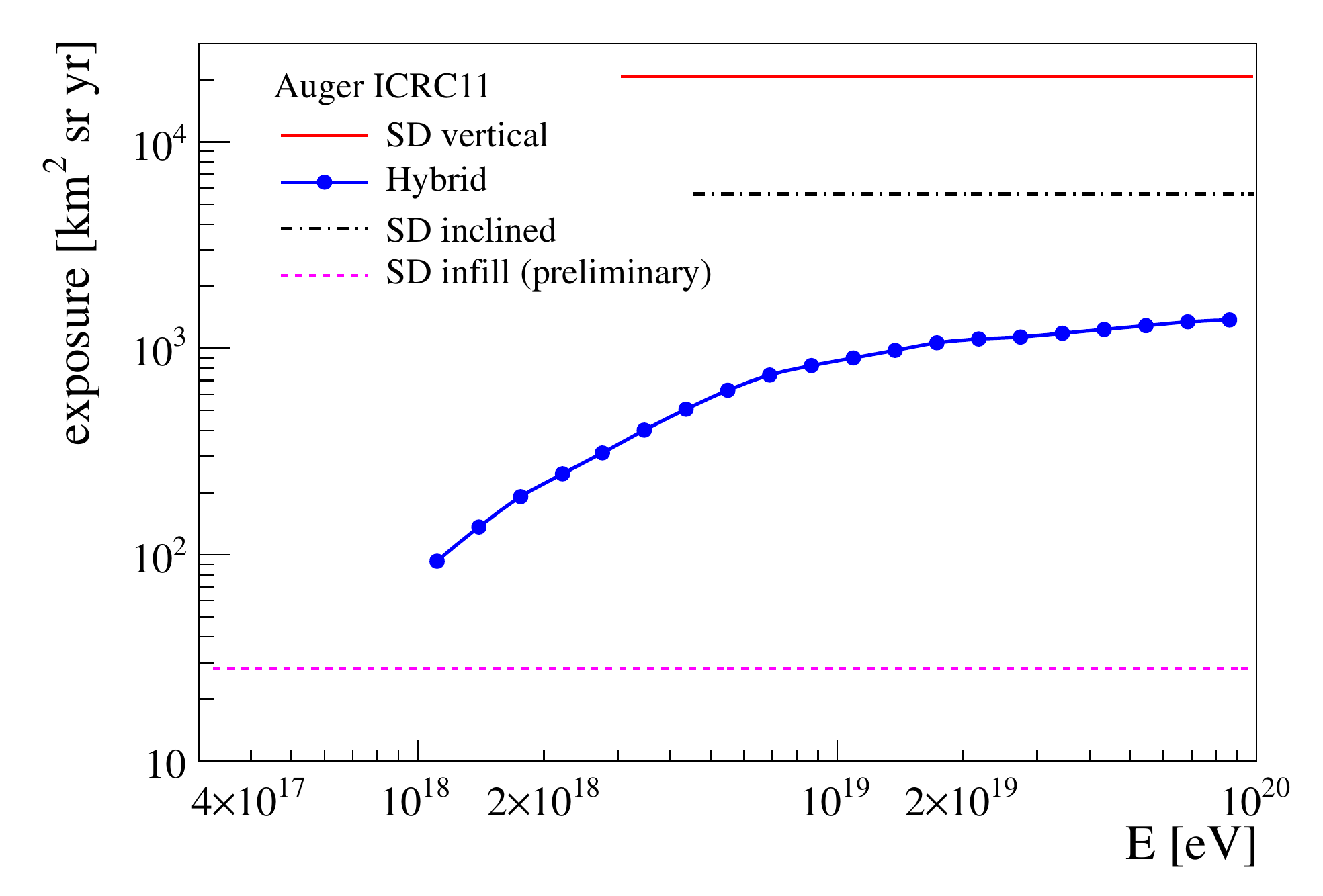}
  }
  \caption{Exposures for the vertical and inclined SD measurements
    together with the hybrid one. The SD exposures are illustrated
    only for the energy range of full trigger efficiencies. The
    exposure for the \unit[750]{m} spacing array (infill) is also shown.}
  \label{fig:exposure}
\end{figure}

\begin{figure}\sidecaption
  \resizebox{0.62\columnwidth}{!}{
    \includegraphics{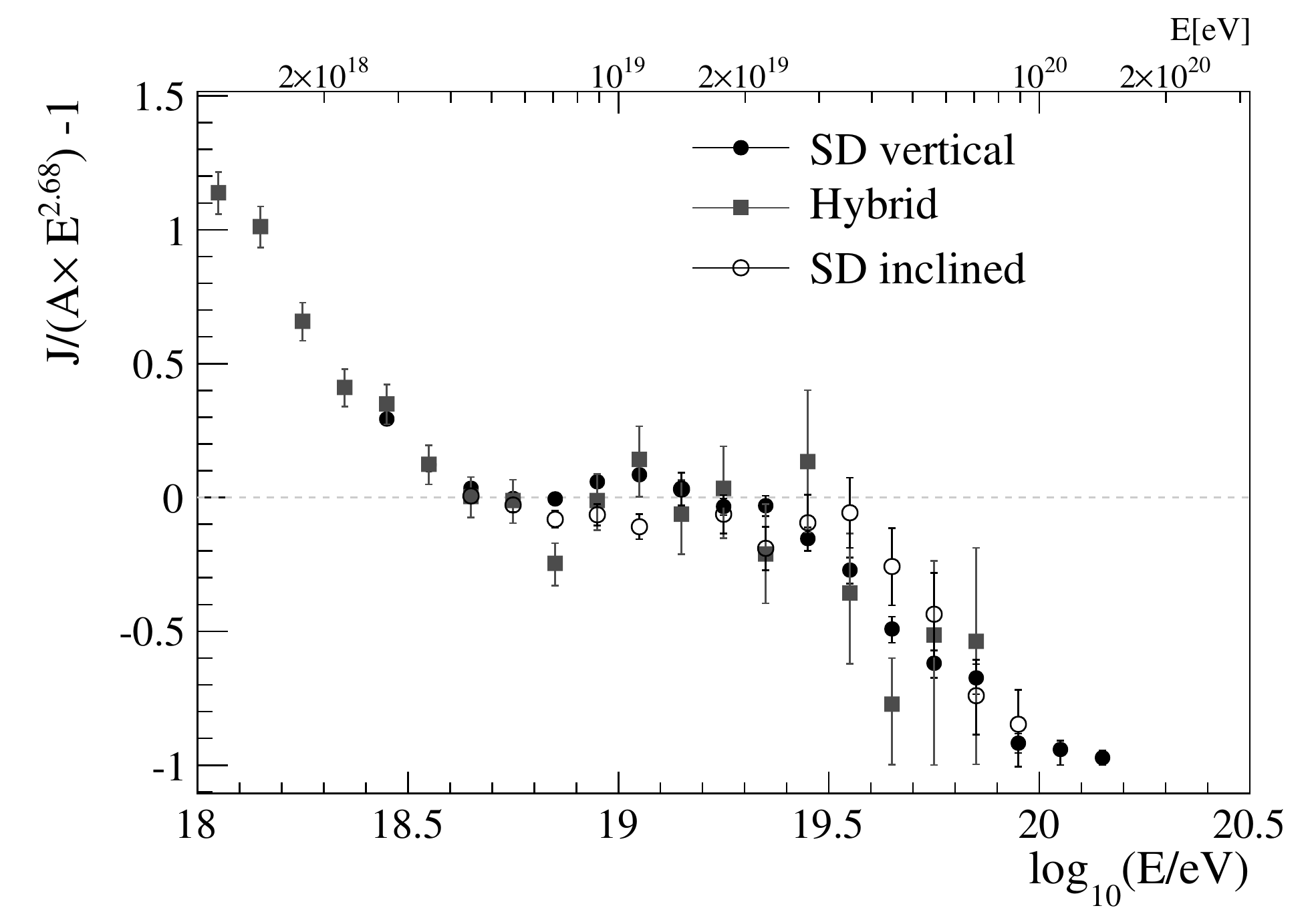}
  }
  \caption{The energy spectra measured with the data recorded at the
    Pierre Auger Observatory. All the three measurements are in a very
    good agreement in the energy range of overlap.}
  \label{fig:enSpectrum}
\end{figure}

\begin{figure}\sidecaption
  \resizebox{0.6\columnwidth}{!}{
    \includegraphics{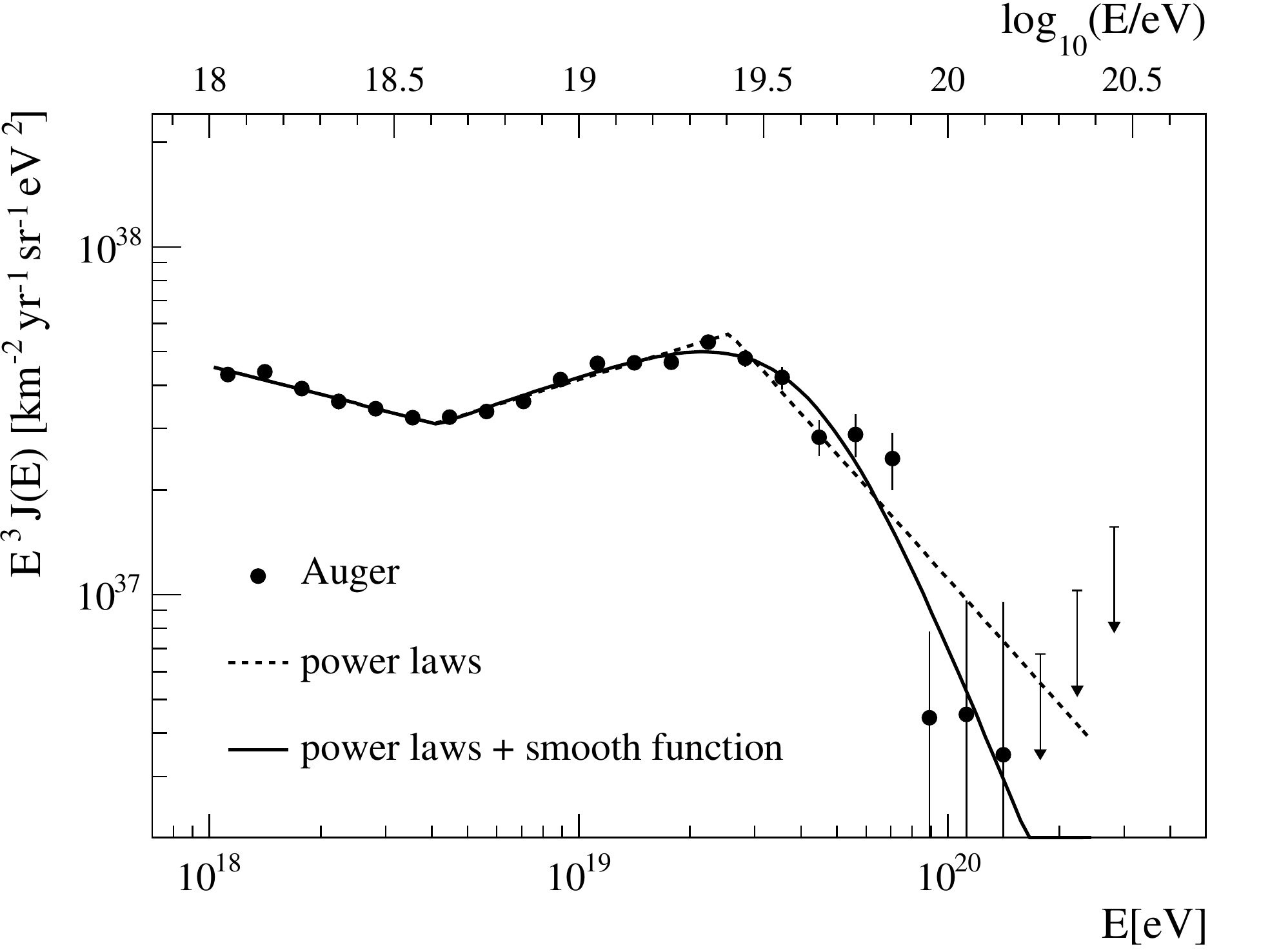}
  }
  \caption{The features of the Auger energy spectrum as obtained from
    fits with two functions (see text). The systematic uncertainties
    on the energy mount up to 22\% and are not shown in the graph.}
  \label{fig:enSpectrumFit}
\end{figure}

The exposure calculation for the surface detector is based on
the time integration of the instantaneous effective
area~\cite{SDAperture}. It amounts up to end of 2010 to
\unit[20905]{km$^2$ sr year} for the vertical data, while for the
inclined events it is four times smaller due to the reduced solid
angle considered (zenith angle between 62 and 80 degrees). The SD
energy spectra~\cite{RobertoICRC,HansICRC} are deduced only in the
energy range with full trigger efficiency, i.e. $\unit[3\times
  10^{18}]{eV}$ for the vertical events and $\unit[4\times
  10^{18}]{eV}$ for the inclined events. The enhancement of the
surface detector with a denser array~\cite{AMIGAICRC} (\unit[750]{m}
spacing between detectors) has been completed. It will increase the
energy range for the SD by one decade. The integrated exposure for the
same period amounts to \unit[29]{km$^2$ sr year}~\cite{IoanaICRC}.
In case of the hybrid measurement the exposure calculation is based on
detailed simulations of the detector~\cite{FDExposure}, taking into
account its time evolution. The simulations consider, besides the SD
and FD hardware status every 10 minutes, all  the atmosphere
measurements and monitoring data. The lower energy limit is given by
the trigger probability of the SD: above \unit[$10^{18}$]{eV} all
events with a zenith angle of less than 60 degrees produce a signal in
at least one SD station~\cite{ltp}. The exposures are represented
in~Fig.~\ref{fig:exposure}, while the energy spectra obtained with the
three data sets are illustrated in Fig.~\ref{fig:enSpectrum}. The
agreement between the energy spectra is very good, the difference in
fluxes is less than 5\%. All the energy spectra share the same
systematic uncertainty on the energy, 22\%, due to the
cross-calibration of the surface detector data with the hybrid energy
determination.  The systematic uncertainties caused by the exposure
determination and energy resolution effects, are independent.  For the
hybrid measurement they vary from 10\% at \unit[$10^{18}$]{eV} to 6\%
at \unit[$>10^{19}$]{eV}. In the case of the vertical surface
detector they are 6\% (3\% exposure determination and 5\%
from the unfolding of the energy resolution effects) while for the
inclined spectrum the exposure contributes with a similar systematic
uncertainties.
Taking into account the independent systematic uncertainties of the
hybrid and vertical SD measurements we obtain an energy spectrum which
extends over the whole energy range above \unit[$10^{18}$]{eV}
(Fig.~\ref{fig:enSpectrumFit}).  The inclined spectrum has not been
yet deconvoluted for the energy resolution effects and thus not
included in the combination. To determine the changes in the flux
evolution with energy we fit two functions: three power laws with free
breaks and two power laws with a sigmoid suppression. The spectral
index changes from $(-3.27\pm0.02)$ to $(-2.68\pm 0.01)$ at
$\log_{10}(E/{\rm eV}) = 18.61\pm 0.01$ and then to $(-4.2\pm0.1)$ at
$\log_{10}(E/{\rm eV}) = 19.41\pm 0.02$.

\section{Conclusions}
\label{sec:conc}
The flux of cosmic rays with energies above \unit[$
  10^{18}$]{eV} has been measured with the data of the Pierre Auger
Observatory, with three independent methods. The agreement between the energy
spectra in the regions of overlap is better than 5\%, well
within the independent systematic uncertainties. The change in the
spectral index at \unit[$(4.1\pm 0.2)\times 10^{18}$]{eV}, the ankle,
and the flux suppression starting at \unit[$(2.6\pm 0.2)\times
  10^{19}$]{eV} have been measured with {very large statistics}. The
current measurements are dominated by the systematic uncertainties on
the energy, 22\%. To achieve a more accurate measurement of the
spectral features the systematic uncertainties on the energy, mostly
due to the fluorescence yield and reconstruction, will be reduced.
The extension of the array with the denser sub-detector will allow the
measurement of the spectral slope down to \unit[$4\times 10^{17}$]{eV}
and the overlap with the KASCADE-Grande measurement~\cite{KG}. A
further enhancement to even lower energies, with an array of
\unit[350]{m} spacing, is foreseen.  We have measured 77 events above
\unit[$5\times 10^{19}$]{eV} in the vertical surface detector data
set, corresponding to about 4 years of full detector efficiency. In
order to accumulate more statistics at the highest energies, a
continuation of the data taking and of the detector maintenance is
essential.

%
%
%

\end{document}